# Integrating En Route and Home Proximity in EV Charging Accessibility: A Spatial Analysis in the Washington Metropolitan Area

**Asal Mehditabrizi**
Graduate Research Assistant,
Maryland Transportation Institute (MTI), Department of Civil and Environmental
3105 Jeong H. Kim Engineering Building, University of Maryland, College Park MD 20742, USA.
Email: asal97@umd.edu

**Behnam Tahmasbi**
Graduate Research Assistant,
Maryland Transportation Institute (MTI), Department of Civil and Environmental
3105 Jeong H. Kim Engineering Building, University of Maryland, College Park MD 20742, USA.
Email: behnamt@umd.edu

**Saeed Saleh Namadi**
Graduate Research Assistant,
Maryland Transportation Institute (MTI), Department of Civil and Environmental
3105 Jeong H. Kim Engineering Building, University of Maryland, College Park MD 20742, USA.
Email: saeed@umd.edu

**Cinzia Cirillo**
Professor
Maryland Transportation Institute (MTI), Department of Civil and Environmental
3250 Jeong H. Kim Engineering Building, University of Maryland, College Park MD 20742, USA.
Email: ccirillo@umd.edu

**Abstract**

This study evaluates the accessibility of public EV charging stations in the Washington metropolitan area using a comprehensive measure that accounts for both destination-based and en route charging opportunities. By incorporating the full spectrum of daily travel patterns into the accessibility evaluation, our methodology offers a more realistic measure of charging opportunities than destination-based methods that prioritize proximity to residential locations. Results from spatial autocorrelation analysis indicate that conventional accessibility assessments often overestimate the availability of infrastructure in central urban areas and underestimate it in peripheral commuting zones, potentially leading to misallocated resources. By highlighting significant clusters of high-access and low-access areas, our approach identifies spatial inequalities in infrastructure distribution and provides insights into areas requiring targeted interventions. This study underscores the importance of incorporating daily mobility patterns into urban planning to ensure equitable access to EV charging infrastructure and suggests a framework that other regions could adopt to enhance sustainable transportation networks and support equitable urban development.

## 1. Introduction

Human mobility poses considerable challenges for both individuals and societies. Motorized vehicles, primarily those with combustion engines, significantly contribute to global warming through greenhouse gas emissions, pose health risks to citizens, and increase global dependence on nonrenewable fuels (1). One emerging solution to these issues is transitioning from gasoline to electricity as the primary transportation fuel, aiming to reduce fossil fuel reliance and mitigate localized air pollution and traffic noise in urban areas. Despite government subsidies for electric vehicles (EVs), there are still substantial barriers to their widespread adoption (2). Key obstacles include the high upfront purchase cost, limited travel range, range anxiety, long charging times, and inadequate charging infrastructure (3–5). However, with EVs projected to reach price parity with conventional vehicles within five to ten years and their ranges now comparable to those of traditional vehicles (6,7), the primary challenge remains the insufficient charging infrastructure (8).

The availability and affordability of charging infrastructure are crucial for the widespread adoption of EVs from multiple perspectives. While home chargers are predominant among EV users, public charging stations are increasingly vital for enabling consistent EV use among apartment residents, renters, and others without access to home chargers (9–11). Adequate public charging infrastructure not only reduces range anxiety and facilitates long-distance travel, thus motivating consumer adoption of EVs (12), but also plays a pivotal role in urban planning by stimulating investments from both private and public sectors in EV charging infrastructure (13). This, in turn, promotes growth in the EV market. On the other side, as the demand for EVs continues to rise, the urgent development of public charging stations has become a global concern to meet this increasing need (14,15).

The development and convenience of public EV charging stations can be effectively assessed through accessibility measures. Accessibility is a well-established method for evaluating the interaction between land use and transportation systems, reflecting how easily individuals can reach specific activities or destinations (16,17). Spatial accessibility, a key component in this analysis, considers factors such as demand, facility supply, and spatial impedance between demand points and facilities (17). Several methodologies are employed to measure accessibility, each offering unique insights. Among these approaches, gravity-based and cumulative opportunity-based methods have been widely used in various studies to evaluate access to facilities such as hospitals, shopping centers, schools, and transportation hubs (18). The gravity-based approach aggregates opportunities or attractions within a study area, applying a discount based on increasing time and distance from the origin (19,20). On the other hand, the cumulative opportunity approach assesses the total attractions or opportunities within a defined distance or travel time from the origin (21,22).

Recently, various methods for calculating accessibility to public services have been applied to evaluate the accessibility of EV charging stations. Specifically, gravity-based and opportunity-based approaches, along with other statistical techniques, are frequently used in this context (23–25). These studies can be categorized into three main groups based on their accessibility calculation methods. The first group employs simple statistical and mathematical techniques, calculating accessibility by determining either the probability of having at least one public EV charging station within a specific area or by counting the number of charging stations available in the region (12,13). The second group uses the gravity-based approach, which incorporates competition among EV charging stations into the model (23,25). The third group applies the opportunity-based approach, predominantly using Gaussian-based Two-Step Floating Catchment Area (2SFCA) methods to assess the accessibility of EV chargers (26–28). Among these approaches, the cumulative opportunity-based method is the most widely used in planning and policymaking due to its comprehensibility and ease of interpretation. In contrast, other methods, such as gravity-based approaches, may overestimate accessibility in isolated areas and involve more complex computations (29,30).

One of the most important direct applications of accessibility measurement to urban facilities is assessing how services are equitably distributed spatially. Previous accessibility measures have been used to evaluate this by analyzing the equity in calculated accessibility. Various methods have been employed to assess equity based on accessibility, including Gini coefficients (27), regression models (23), and spatial autocorrelation analysis (25). Among these, spatial analysis is particularly powerful for examining the equitable distribution of urban services because it simultaneously incorporates location and attribute information (31). This technique is widely used in studies related to EV infrastructure. Research employing spatial analysis has been conducted in diverse geographic locations, including China, California, and Washington State. These studies consistently reveal that public EV charging stations are not distributed equitably, with lower-income communities experiencing reduced accessibility to these facilities (12,13,23,25). This highlights ongoing disparities in the availability of EV charging infrastructure and underscores the need for targeted policies to address these inequities.

The inequitable distribution of EV charging infrastructure across different states in the U.S. and the crucial role of accessibility in equity analysis emphasize the need for an appropriate method for calculating accessibility that offers a realistic measurement. Previous studies on accessibility to EV charging stations often employ methods similar to those used for other public services. Although these studies attempt to enhance realism by considering factors such as competition among charging stations and the demand for EVs, they generally overlook the significant differences in drivers' approaches to reaching charging stations compared to other public facilities. Specifically, drivers utilize two main types of charging: destination charging and en route charging (32). Destination charging occurs when EV charging is a supplementary service at locations like hotels, supermarkets, and gyms, and drivers' decisions to park are independent of the vehicle's charge level (33–36). In contrast, en route charging involves drivers charging their vehicles along their travel route, influenced primarily by the vehicle's charge state and range anxiety (14,37–41). Previous accessibility measurements have predominantly focused on destination charging, while en route charging is also vital. It not only increases the number of charging opportunities for drivers but also affects accessibility by increasing competition for charging stations. Thus, incorporating en route charging into accessibility assessments could provide a more comprehensive understanding of charging infrastructure needs and equity.

A significant gap in previous studies on the accessibility and equity of EV charging stations lies in both methodological perspectives and geographic focus. Most research has concentrated on the West Coast of the United States, particularly in California and Washington State, with no comprehensive investigation into the Washington metropolitan area. Additionally, the methodologies used often fail to consider the unique dynamics of travel to charging stations. Accessibility is a crucial lens through which to examine the interplay between land use and transportation systems, especially for facilities like EV charging stations. Previous research has frequently applied models designed for measuring accessibility to general amenities such as hospitals, workplaces, and parks, without fully considering that trips to a charging station might not only serve as the primary activity but could also be integrated as a detour en route to other destinations, such as workplaces. Therefore, it is imperative to first understand the underlying logic of this interaction between land use and transportation for charging stations. Developing a tailored model that accurately captures and measures accessibility in this context is essential to address these complex travel behaviors effectively. To address these limitations and better understand the validity of previous findings, it is crucial to develop and apply a more comprehensive measurement of accessibility that incorporates both destination and en route charging. Comparing this enhanced measurement with previous destination-based methods will provide a more accurate assessment of the equity of EV charging infrastructure.

This study makes several innovative contributions to the existing body of research. Firstly, it introduces a more realistic methodology for measuring accessibility by incorporating both destination and en route charging. This new approach is a modified version of the traditional opportunity-based method and includes factors such as competition for EV charging stations and varying capacities at each station. Secondly, the study calculates accessibility using both the new method and previously established approaches. By comparing these methods within the study area, the research identifies where the new methodology provides significant improvements over destination-based measures. Finally, the study evaluates spatial equity in the distribution of charging stations, revealing various forms of inequitable distribution based on spatial patterns across different states within the Washington metropolitan area.

## 2. Data

This study utilizes two sets of datasets: the 2017/2018 Regional Household Travel Survey data from the Washington Council of Governments (WSCOG), EV charging station location data from the Department of Energy's (DOE) Alternative Fuels Data Center, and the 2018 American Community Survey data. The Regional Household Travel Survey data is used to calculate the number of trips originating from and arriving at each census tract. The study area is the Washington metropolitan area, which includes parts of District of Columbia, Maryland, Virginia, and West Virginia. This dataset comprises four subsets: household, person, vehicle, and trip information. The primary focus of this study is the trip subset, which includes data on 126,874 trips made by 32,923 individuals. Each trip is assigned a specific weight, which is used to generate origin-destination (OD) trip information at the census tract level.

The second set of datasets used in the study includes the EV charging stations' location data, which provides information on each station's location (longitude and latitude), type, and the number of plugs available. By merging this data with the OD trip information derived from the regional travel survey, the locations of 1,576 public charging stations in the Washington metropolitan area were identified. These stations have different types of plugs, including Level 2 and DC fast charging. Level 2 charging stations can provide up to 35 or 40 miles of range per hour, making them suitable for short errands within 20 miles of home, such as shopping or dining, and for vehicles parked for more than two hours. In contrast, DC fast charging stations can charge an EV to 80% in just 20 minutes, offering a convenient refueling option for longer trips (42).

## 3. Methodology

In this study, computational methods are utilized for integrating datasets, measuring accessibility, and visualizing spatial analyses. The primary objective is to assess accessibility using two distinct approaches and compare the outcomes through spatial analysis and equity evaluation.

### *3.1. Assessment of Electric Vehicle Charging Station Accessibility*

Accessibility is a critical factor in evaluating the development of public EV charging infrastructure. Therefore, the methodology for measuring accessibility must meet several criteria: it should be based on behavioral considerations, technically feasible, and easy to interpret (43).Research on EV charging stations and spatial equity is relatively sparse (23,25,27). Existing studies primarily use gravity-based and opportunity-based approaches, which have also been applied to other public services such as healthcare, public transportation, groceries, and parks (44,45). These approaches often assume that drivers' behavior when accessing EV charging stations is similar to their behavior when visiting other amenities, such as parks or grocery stores. This assumption overlooks an important consideration: the behavioral basis of accessibility measurement. Driver behavior for charging EVs differs from that for other public amenities.

Drivers use two distinct approaches when accessing EV charging stations as depicted in Figure 1 (32). The first approach involves activities that typically start from home and end at a charging location near the residence. Within this approach, drivers might make a dedicated trip specifically for charging or charge their vehicles while running errands, such as shopping or dining, where the primary purpose of the trip is unrelated to charging. In this approach, the convenience of charging stations near home is a significant factor in their decision. The second approach involves charging the vehicle as a detour while traveling to another destination, where the primary purpose of the trip is not related to charging. In this case, the driver stops to charge their vehicle as a secondary activity during their journey to the final destination.

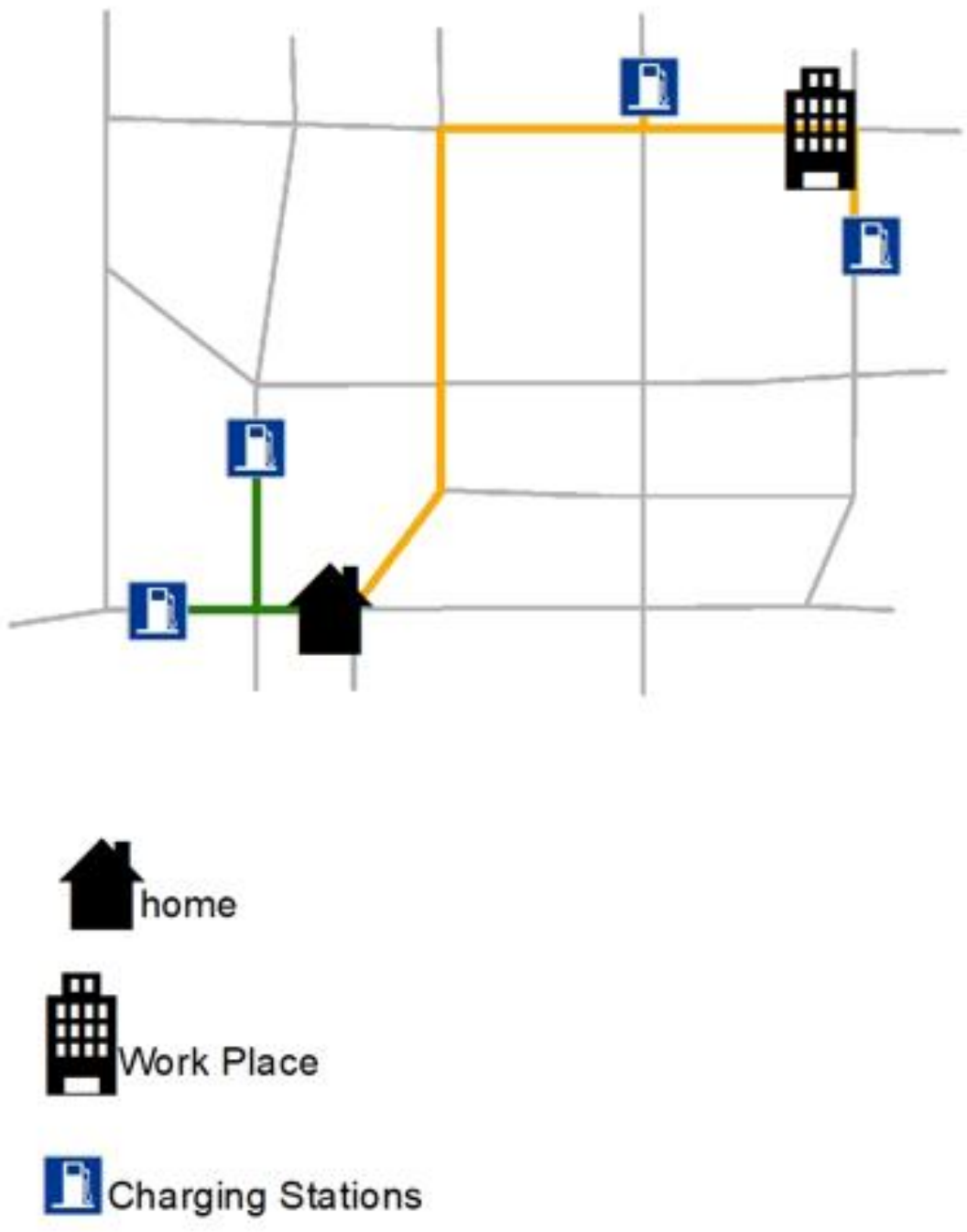

Figure 1. Symbolic representation of the two EV charging approaches: the orange line indicates en route charging; The green line represents destination charging.

Previous studies have predominantly focused on the first approach, neglecting the second approach. To address this gap, this study calculates accessibility in two steps using a more practical and realistic opportunity-based approach (22,25). In the first step, accessibility is calculated for conditions when drivers leave home to charge around their residential area (destination charging). In the second step, accessibility is calculated for charging as a detour along the route to a destination (en route charging). The calculated accessibility is then compared with the accessibility measured by the traditional opportunity-based approach used in previous studies (destination-based approach) (25). The steps for calculating accessibility are as follows:

*3.1.1. Destination charging*

This step addresses scenarios where drivers either leave home specifically to charge their vehicle or charge their vehicle while running errands for other activities. In this context, drivers generally have more time for charging their vehicle compared to charging along a route. Therefore, both level 2 and DC fast charging plugs are considered for accessibility measurement. An opportunity-based model is utilized to determine the accessibility of charging stations. To make the opportunity-based model more realistic and improve its accuracy, it is essential to consider charging demand. The actual accessibility to a public service like EV charging is affected by the level of demand for that service (46). Since the opportunity to access EV charging services is limited and exclusive to a single user, individuals within a service radius must compete to use the charging service. Consequently, a competition factor is included in measuring accessibility. The accessibility formulation is as follows:

$$A1_i = \sum_{k \in d_{ik} < d1_0} \frac{R_k}{d_{ik}^{\beta}} \qquad (1)$$

In this equation, $A1_i$ represents the accessibility of census tract $i$ based on the first step. $d_{ik}$ is the Euclidean distance between the demand location (centroid of census tract $i$) and public EV charging station $k$. $d1_0$ is the predefined threshold of spatial impedance, which is set to 15 miles in this study. This distance is chosen because individuals typically prefer not to travel beyond a certain distance from their residence to charge their vehicle or to shop and dine (26,42). $R_k$ is the relative service availability of charging station k, calculated based on a supply-to-demand ratio, which will be explained comprehensively in the following sections. $\beta$ is the friction coefficient, set to 2 based on previous literature (23).

*3.1.2. En route charging*

Individuals may have access to charging stations within a particular detour from their route between origin and destination. (40) demonstrate that people are willing to deviate up to one mile from their route to reach a station. In this section, only DC fast charging stations are considered, as the primary purpose of the trip is not to charge the vehicle; therefore, charging should be completed in the shortest possible time. The opportunity-based function utilized in this scenario is depicted in Equation 2.

$$A2_i = \sum_{j \in J} \alpha_{ij} \sum_{k \in\, d_{ikj} < d2_0} \frac{R_k}{d_{ikj}^{\beta}} \tag{2}$$

In this equation, $A2_i$ is the accessibility of census tract $i$ based on the second step. $d_{ikj}$ is the detour from a route from origin $i$ to destination $j$ to reach station $k$. $d_0$ is the predefined threshold for detour which is set to one mile (40). $R_k$ is the relative service availability of charging station $k$, the same as in the first step. $\alpha_{ij}$ is the percentage of trips originating from census tract $i$ and destined for census tract $j$. In other words, $\alpha_{ij}$ is a weight given to the accessibility of each route originating from zone $i$, indicating the importance of each origin-destination pair's accessibility to charging stations based on the percentage of trips traveling between this pair.

The final accessibility index is obtained by summing the accessibility values from these two steps ($A = A1_i + A2_i$).

*3.1.3. Relative Service Availability*

There is competition among communities to access EV charging services because the number of reachable charging opportunities for a community decreases when the station is also accessible to other communities. In this study, a relative approach is proposed to approximate the share of accessible opportunities within a threshold and average it based on the total charging demand of each community. To achieve this, it is essential to understand the service availability of each charging station, as the service capacity at each station depends on the types of charging facilities, including DC fast and Level 2 charging plugs. DC fast chargers have higher charging speeds and nominal power compared to Level 2 chargers (32); therefore, stations with more DC fast chargers can serve more users than stations with a greater number of Level 2 chargers.

In this study, the relative serviceability is divided into two parts to account for the different charging types, as their supply (serviceability) as well as demand for them varies based on distinct charging approaches (destination charging vs. en route charging). Since DC fast charging plugs are available to both people charging near their homes and those charging along their routes, the demand for a station with DC fast plugs includes demand from communities within a 15-mile threshold of the station and demand from trips where the station is a one-mile detour from the route. For Level 2 chargers, the demand comes only from communities within the 15-mile threshold of the station. The resulting formulation for the relative serviceability is as follows:

$$R_k = \frac{\gamma N_{DC_fast}}{\sum_{i \in \{d_{ik} < d1_0\}} \frac{D1_k}{d_{ik}^{\beta}} + \sum_{ij \in \{d_{ikj} < d2_0\}} \frac{D2_k}{d_{ikj}^{\beta}}} + \frac{N_{level2}}{\sum_{i \in d_{ik} < d1_0} \frac{D1_k}{d_{ik}^{\beta}}} \tag{3}$$

In this equation, $R_k$ represents the serviceability of station $k$. $N_{DC_fast}$ and $N_{k2}$ denote the number of DC fast chargers and Level 2 chargers at station $k$ respectively. The parameter $\gamma$ reflects the relative daily maximum number of vehicles that a DC fast charger can service compared to a Level 2 charger; this is set to 6, based on the fact that the average nominal power of a DC fast charger is approximately 6 times greater than that of a Level 2 charger (Esmaili et al., 2024). $D1$ represents the demand at origin K, based on the population originating from census tracts within a 15-mile buffer of station $k$. $D2$ represents demand based on trips between tract $i$ and tract $j$ where station $k$ is within a one-mile detour. All other variables are as described in equations 1 and 2. The dominators of the first and the second part represent the competition impact for DC fast charging stations and level 2 charging stations, respectively. The competition factor formulation incorporates distance to reflect the reduced probability of simultaneous high demand over a larger area.

*3.2. Destination based Methodology for Assessing Accessibility to Electric Vehicle Charging Stations*

In this study, accessibility is also measured using an opportunity-based method similar to that employed in previous studies(25), as outlined in Equation 1. The primary distinction from the previous approach is the competition factor, which in this case is solely based on the 15-mile threshold from each station. The formulation for this method is as follows:

$$R_k = \frac{\gamma N_{DC_fast} + N_{level2}}{\sum_{i \in \{d_{ik} < d1_0\}} \frac{D1_k}{d_{ik}^{\beta}}} \tag{4}$$

All of the variables in this equation are the same as those in Equation 3.

*3.3. Spatial equity of EV Charging Stations distribution*

This study employs spatial autocorrelation analysis to investigate the equity in the distribution of EV charging stations, focusing on the spatial patterns of our accessibility measure. Both Global and Local Moran's I statistics are used to assess spatial autocorrelation (47). Moran's I is a spatial statistics measure that reveals the extent of variation between attributes at nearby geographic locations, thereby indicating the level of equitable distribution of public services such as EV charging stations (31). The measure ranges from 1 to -1, where positive values indicate that neighboring areas have similar levels of accessibility, and negative values suggest that nearby areas have differing accessibilities. A value near zero, or one that is statistically insignificant, indicates a random spatial distribution (48,49), suggesting that the distribution of EV charging stations is fairly equitable. The Global Moran's I measures spatial autocorrelation over the entire study area, and is calculated as follows:

$$I = \frac{M}{W} \frac{\sum_{i=1}^{M} \sum_{j=1}^{M} w_{ij} (A_i - \bar{A})(A_j - \bar{A})}{\sum_{i=1}^{M} (A_i - \bar{A})^2} \tag{5}$$

In this equation, $I$ represents the Global Moran's I statistic. The term $M$ denotes the total number of census tracts, indexed by $i$ and $j$. $A$ refers to the measured accessibility, while $\bar{A}$ is the average accessibility across all tracts. The elements $w_{ij}$ form a spatial weights matrix, with diagonal elements ($w_{ii}$) set to zero. For constructing the weight matrix, this study uses the K-nearest neighbor (KNN) method.

Specifically, the 10 nearest neighbors for each census tract $i$ are identified, and $w_{ij}$ is assigned a value of 1 if census tract $j$ is among these neighbors, and zero otherwise. $W$ is the sum of all $w_{ij}$ values, computed as follows:

$$W = \sum_{i=1}^{M} \sum_{j=1}^{M} w_{ij} \tag{6}$$

While the Global Moran's I measures spatial autocorrelation over the entire region, the Local Moran's I focuses on the spatial pattern surrounding each specific observation (50). As a variant of the Local Indicator of Spatial Association (LISA), Local Moran's I identifies significant spatial patterns and highlights clusters and outliers within individual geographic areas. The formula for Local Moran's I is given by:

$$I_i = \frac{(A_i - \bar{A})}{S^2} \sum_{j=1}^{M} w_{ij}(A_j - \bar{A}) \tag{7}$$

In this equation, $S^2$ is determined as follows:

$$S^2 = \frac{\sum_{i=1}^{M}(A_i - \bar{A})^2}{M} \tag{8}$$

All other variables referenced in Equation (8) have been previously defined in the context.

We compute the Local Moran's I statistic for each census tract to assess the statistical significance of each $I_i$, which helps in identifying various spatial patterns within residential communities (50). This analysis can reveal four distinct types of spatial configurations: high-high clusters, low-low clusters, high-low outliers, and low-high outliers. A high-high cluster emerges when a community with high accessibility to EV charging stations is surrounded by other areas with similarly high accessibility. Conversely, a low-low cluster occurs when a community with low accessibility is encircled by areas of low accessibility. Significant negative values highlight considerable disparities between a community and its neighboring areas, resulting in high-low or low-high outliers. If the values are not statistically significant, it indicates that the accessibility levels of a community are not strongly correlated with those of its neighboring areas.

## 4. Results

### *4.1. Charging accessibility measurement*

This study measures accessibility in the Washington metropolitan area using two different methods. In this section, the results of the proposed method are presented and compared with those of destination-based methods, which only consider trips that originate from home to reach charging stations. To facilitate comparison, the accessibility measures in both methods are normalized. Figure 2 illustrates the normalized accessibility for each census tract calculated by these methods, along with the relative change between them. The comparison between Figures 2a and 2b demonstrates that our method for calculating accessibility reduces the disparity between high commuting areas and core areas observed with the destination-based method. In other words, the number of areas with extremely low or high accessibility decreases, indicating that actual accessibility is more evenly distributed than previously calculated. This improvement is due to the destination-based method only considering the number of stations within or near each tract, resulting in lower accessibility in areas with fewer stations (23,25). This is evident from the higher correlation between the number of charging stations and accessibility in the destination-based method ($p < 0.001$, Pearson

correlation = 0.612) compared to our method ($p < 0.001$, Pearson correlation = 0.541). However, in reality, people do not solely charge their vehicles near home; their daily trips to work, shopping, or other destinations further from their residence also impact their charging opportunities, as they might charge their vehicles along these routes (32). Therefore, our method provides a more accurate reflection of actual accessibility by accounting for these additional charging opportunities, resulting in a more uniform distribution of accessibility across different areas.

Figure 2c shows the relative differences in accessibility calculated by the two methods. It reveals that accessibility in central areas is higher in the destination-based method than in the new one, while the opposite is true for marginal areas. By categorizing these two groups of areas and performing F-statistics, significant differences can be observed in population, demand, and the total weighted number of plugged-in chargers in each census tract. Table 1 indicates that in areas where the accessibility measured by the new method is lower than the previous one, the average weighted number of chargers in each census tract and the average demand are higher. In these areas, the destination-based method shows higher accessibility due to the greater number of chargers, despite the high demand for them. This demonstrates that the positive effect of a high number of charging stations outweighs the negative effect of higher demand, which leads to competition for chargers.

Table 1 reveals that in areas where the destination-based method indicates higher accessibility, the increased number of chargers is largely attributed to the prevalence of DC fast chargers. The destination-based method does not account for en route charging, assuming that drivers compete for both level 2 and DC fast chargers within a fixed distance from their homes. In contrast, the new method shows that competition for DC fast chargers is more intense because they are also used for en route charging along travel routes. As a result, despite the high number of chargers in these areas, the substantial demand for them reduces overall accessibility. Therefore, the accessibility figures provided by the destination-based method in these areas are likely overestimated.

There are also areas where the destination-based method underestimates accessibility. Although these areas have a lower average number of chargers per census tract, this factor alone does not provide a complete picture. The number of chargers in surrounding areas, which offers additional charging opportunities along travel routes, and the number of trips made outside the census tract are crucial for accurately assessing accessibility. Underestimating accessibility in these areas could lead to misallocation of resources, directing them to places where they are less needed. Conversely, overestimating accessibility may result in insufficient focus on areas that require infrastructure improvements, potentially leading to underserved zones.

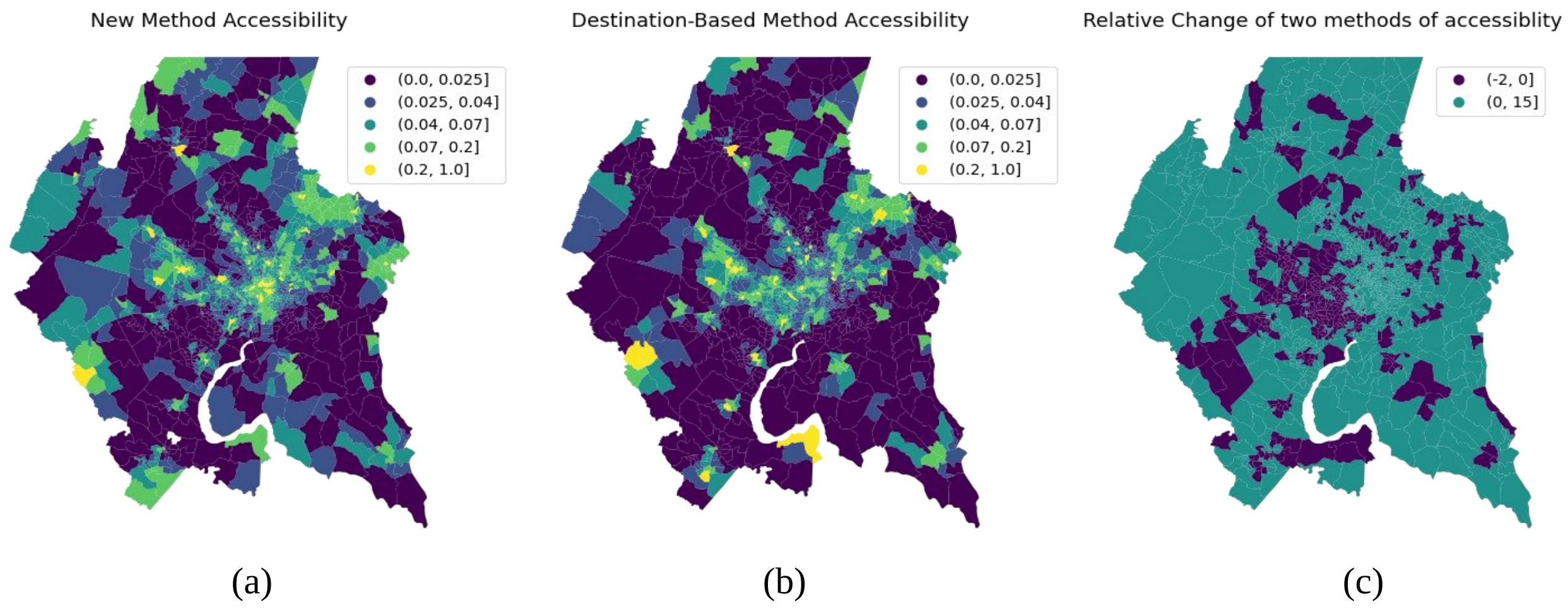

Figure 2. Normalized accessibility by census tract: (a) New method, (b) Destination-based method, (c) Relative changes.

Table 1. The result of F-statistics for comparison between variables in two groups of areas

| Average Variables in Census Tract | Group 1 (New Method Accessibility > Destination-based Method Accessibility) | Group 2 (New Method Accessibility < Destination-based Method Accessibility) | F statistics (p value) |
|---|---|---|---|
| Accessibility (New Method) | 0.067 | 0.0571 | 7.542 (<0.001) |
| Accessibility (Destination-based Method) | 0.046 | 0.080 | 71.381 (<0.001) |
| Population | 4412.107 | 4859.829 | 19.500 (<0.001) |
| Demand | 630,607 | 942,965 | 37.155 (<0.001) |
| Total Weighted Plug-in Chargers | 4.040 | 8.832 | 26.747 (<0.001) |
| Number of Level 2 Chargers | 2.84 | 1.445 | 12.213 (<0.001) |
| Number of DC Fast Chargers | 0.2 | 1.231 | 72.109 (<0.001) |

*4.2. Spatial autocorrelation analysis*

This study conducts a spatial autocorrelation analysis to compare the spatial patterns of EV charging station distribution between two methods. The Global Moran's I values for charging accessibility indicate significant positive autocorrelation for both methods, revealing spatial clusters of high-access and low-access areas (destination-based method: 0.21, p-value = 0.001; new method: 0.30, p-value = 0.001). This suggests that the Washington metropolitan area experiences inequitable access to EV charging stations, regardless of the method used.

To further explore the local patterns of spatial autocorrelation, Local Moran's I is calculated for both methods and their relative differences, as shown in Figure 3. These results reinforce the findings of Global Moran's I, highlighting the presence of adjacent census tracts with similar levels of access to charging stations. The figure illustrates that census tracts are clustered in specific areas, with the majority falling into High-High and Low-Low categories, indicating clusters of high and low accessibility, respectively.

Figures 3a, 3b and 3d demonstrate that low-low clustering is predominantly found in areas with either zero or very few plug-in chargers. The comparison between the two methods reveals that both the number of low-low and high-high clusters is higher in the new method, indicating a more significant inequitable distribution of charging infrastructure than previously estimated. Additionally, it is observed that low-low clusters, which require greater attention, decrease in the northern parts of the area while increasing in the southern parts.

Figure 3c illustrates the clustering of relative differences in accessibility between the two methods. Most census tracts are in areas where the accessibility calculated by the new method is higher than by the destination-based method. While the destination-based method underestimates accessibility in these areas, potentially leading to resource allocation where it is not needed, the low-low areas demand more attention. Specifically, areas categorized as low-low in both Figure 3a (new method accessibility) and Figure 3c (relative change) but not in Figure 3b (destination-based method) should be prioritized. In these areas, the destination-based method overestimates accessibility, failing to identify them as low-accessibility zones, whereas the new method correctly identifies their need for improvement.

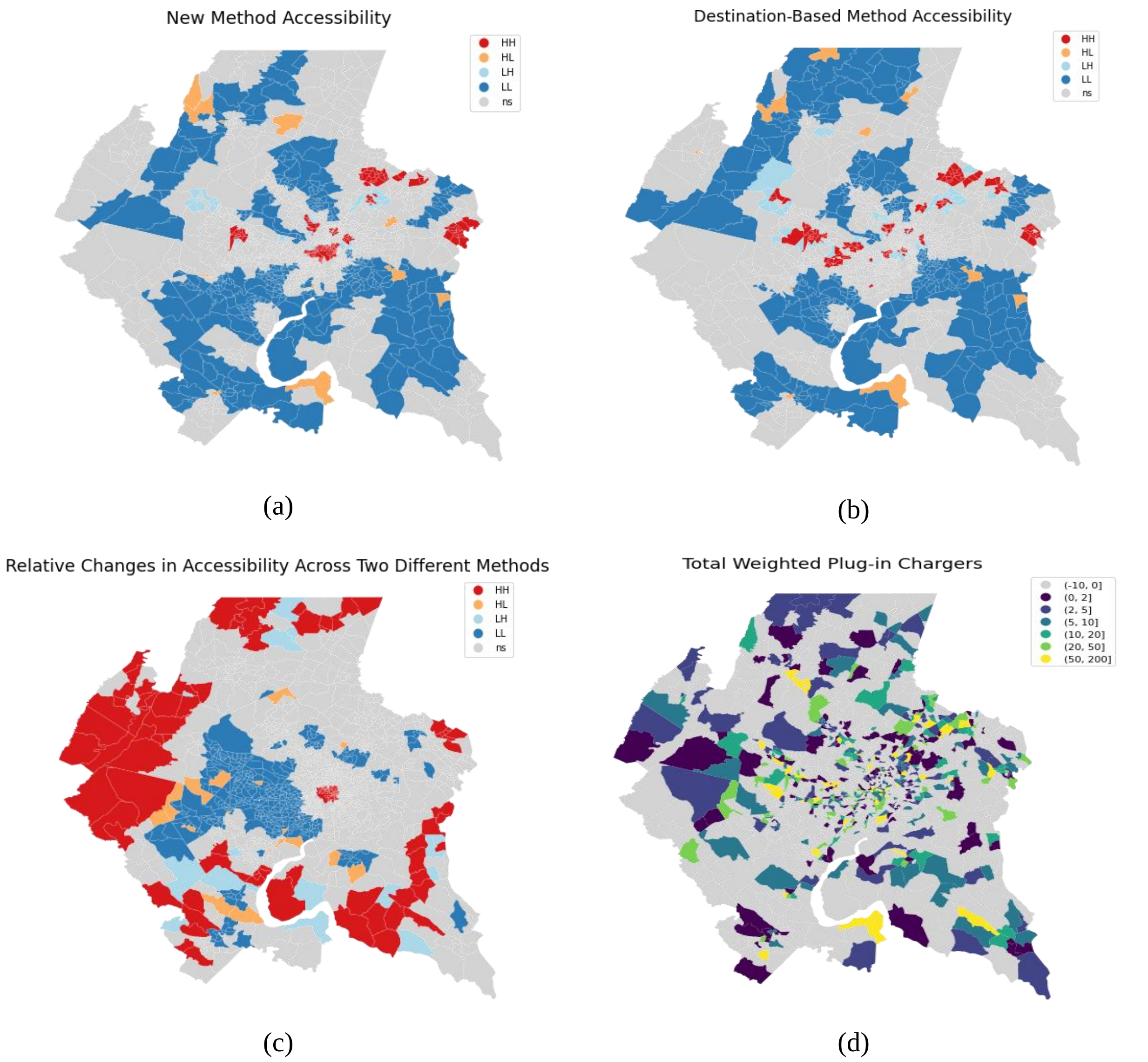

Figure 3. Local Moran's I results: (a) New method; (b) Destination-based method; (c) Relative change in accessibility; (d) weighted number of chargers.

If we divide the Washington metropolitan area into different states as shown in Figures 4, 5, and 6, a notable difference emerges between the two methods. Based on the new method depicted in Figure 4a, 31% of census tracts in the District of Columbia are categorized as high-high (HH), whereas the destination-based method (Figure 4b) categorizes only 7% of census tracts in DC as HH. Conversely, 5% of census tracts in DC are categorized as low-low (LL) according to the new method, compared to 10% with the destination-based method.

Figures 4a and 4c demonstrate that the central part of DC exhibits relatively high level of accessibility, which can be attributed to a greater number of charging stations in the area and its neighboring regions, where people conduct most of their trips. This higher station density helps distribute demand more evenly, reducing competition for both local and en route charging stations.

However, the destination-based method has limitations. It only considers the number of stations within a 15-mile buffer. Consequently, in areas where the number of stations is lower than in neighboring areas, as shown in Figure 4b, accessibility is inaccurately categorized as low. In reality, people can charge their vehicles along their routes, and since neighboring areas have a higher number of stations, the actual accessibility is not as low as the destination-based method estimates.

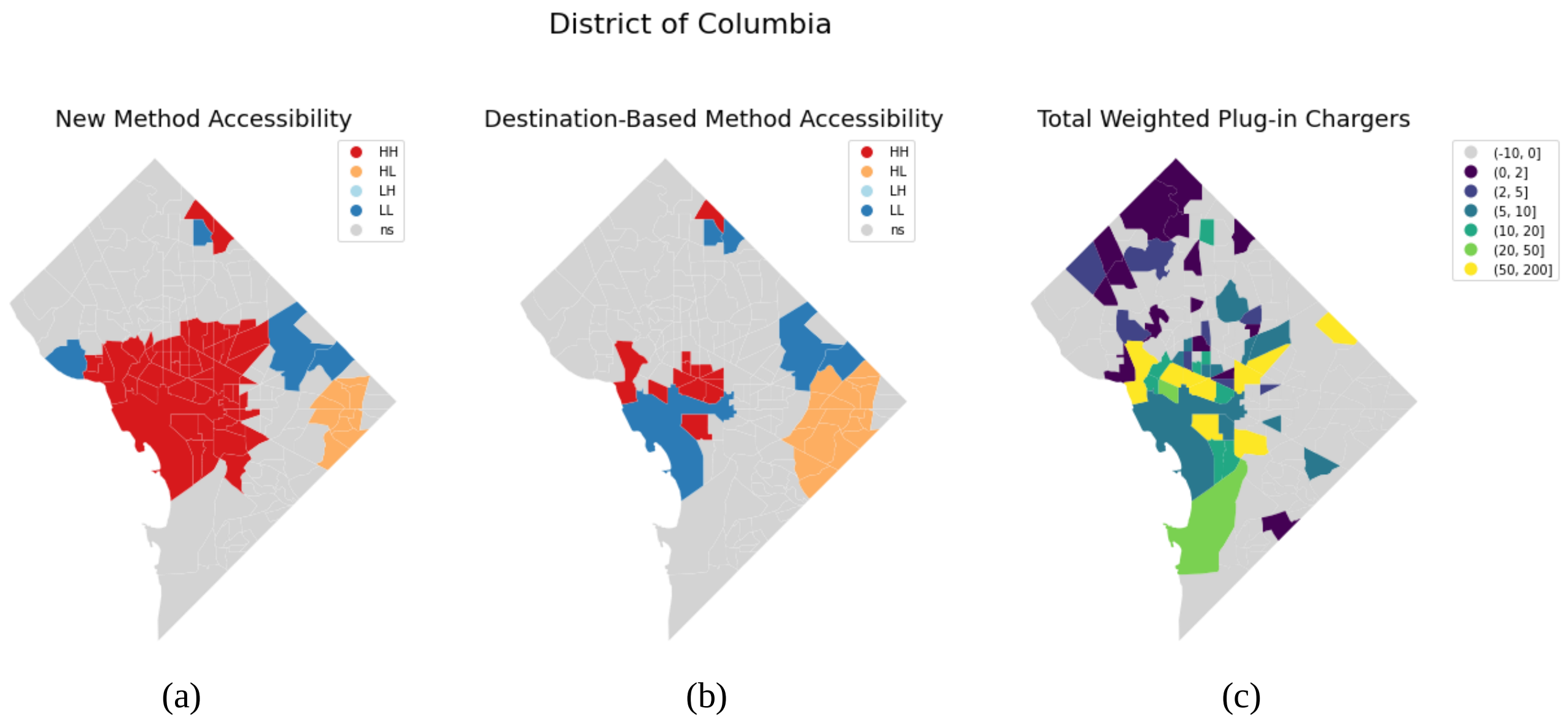

Figure 4. Local Moran's I results for DC: (a) New method; (b) Destination-based method; (c) weighted number of chargers.

Figure 5a and 5b show that a large proportion of census tracts in Virginia are categorized as LL, with 29% according to the new method and 18% according to the previous one. This indicates that Virginia has overall low access to charging stations, a trend supported by the low number of charging stations in many of its census tracts.

The difference between the results of the two methods is smaller in this area because, across much of Virginia, the number of charging stations is low. Consequently, residents have limited access to charging stations, even when traveling from their residential tracts to other locations. This highlights that neighborhood context plays a significant role in calculating accessibility with our method. When both the number of stations in a given area and its neighboring locations are low, the accessibility estimated by our method aligns closely with that of the destination-based method. Conversely, if the number of stations in neighboring areas is higher than in a specific location, our method estimates greater accessibility compared to the destination-based method. This demonstrates that the destination-based method may underestimate accessibility in areas with relatively high numbers of stations nearby.

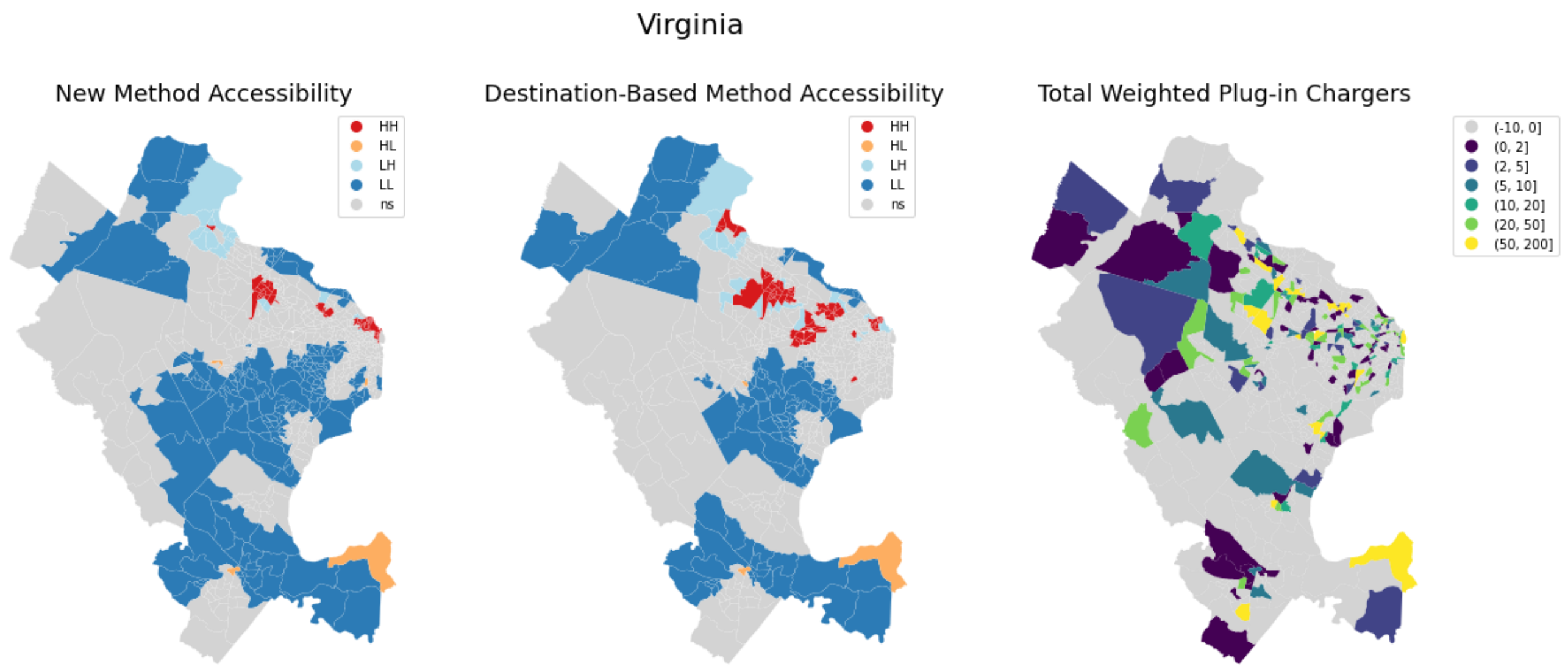

(a) (b) (c)

Figure 5. Local Moran's I results for Virginia: (a) New method; (b) Destination-based method; (c) weighted number of chargers.

Figure 6 illustrates the spatial clustering of accessibility and the number of charging stations in Maryland. According to the new method, 26% of census tracts are categorized as low-low, compared to 29% using the previous method. Conversely, 10% of census tracts are classified as high-high based on the new method, while 8% are categorized similarly by the destination-based method. Overall, these figures indicate that accessibility in Maryland is low and requires further improvement.

Notably, even in census tracts with a high number of charging stations, accessibility can still be assessed as low. This occurs because the high number of trips in the area increases demand and competition for charging stations. Additionally, the location of charging stations significantly affects the actual charging opportunities for residents. Thus, the number of stations in a census tract alone, without considering nearby opportunities and the trip destinations of residents, can lead to overestimated or underestimated accessibility calculations.

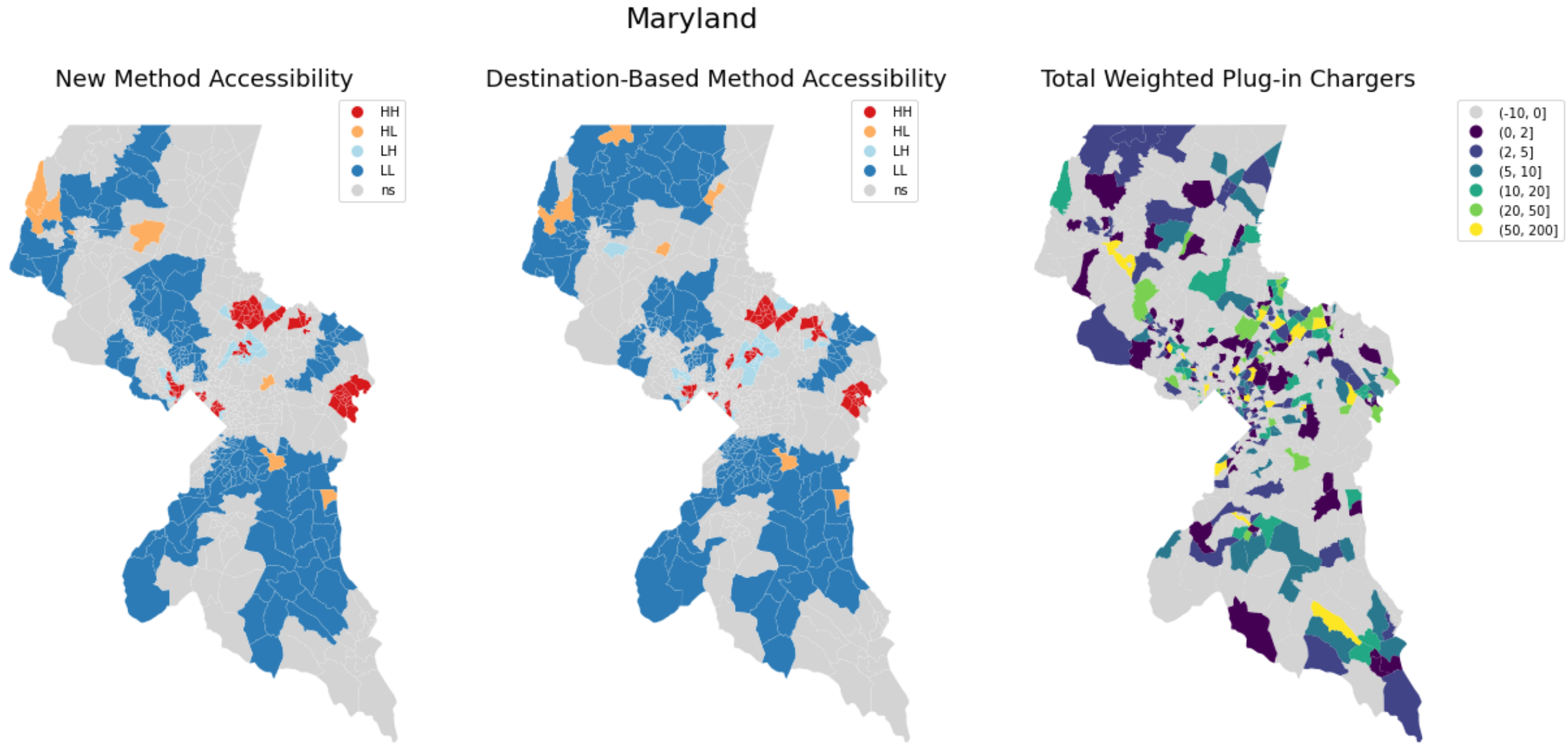

(a) (b) (c)

Figure 6. Local Moran's I results for Maryland: (a) New method; (b) Destination-based method; (c) weighted number of chargers.

## 5. Discussion and Conclusion

This study presents a novel approach to measuring the accessibility of EV charging stations by considering not only their proximity to residential areas but also their distribution along typical travel routes. This method reflects a more comprehensive understanding of accessibility, recognizing that charging behavior is influenced by an individual's entire travel pattern rather than just their immediate residential environment. Significantly, the study highlights that destination-based methods, which focus primarily on proximity to home, tend to overestimate accessibility in core areas while underestimating it in peripheral regions. This discrepancy has led to potential misallocations of resources, where charging infrastructure may not align with actual demand. The results from our analysis suggest that previous assessments might have misdirected resources by overemphasizing central areas while neglecting underserved or commuter-heavy regions. By addressing these inaccuracies, the new method directs attention to areas that are inadequately served, suggesting a more strategic placement of charging stations that accounts for both residential and en route charging needs. This approach has significant implications for urban planning and the equitable distribution of charging infrastructure, highlighting the need for a more nuanced strategy to ensure that resources are allocated effectively.

Moreover, the findings of spatial autocorrelation analysis, showing significant clustering of high-access and low-access areas, underscore the spatial inequality in current infrastructure distribution. The new method's capacity to identify these clusters more accurately points to its utility in informing targeted interventions that address these disparities. For instance, the recognition of low-low clusters in areas previously considered adequately served by charging stations highlights the need for a refined strategy in infrastructure placement that transcends simplistic proximity-based models.

Our study has several implications for policy-making and future research. First, it suggests that urban planning and infrastructure development should consider the full scope of daily travel behavior to ensure that EV charging stations are accessible to all users across their entire travel network. This approach can help mitigate the current inequities in access to charging infrastructure, partly due to relying solely on residential proximity in accessibility assessments.

Second, the methodology presented here can serve as a framework for other metropolitan areas looking to improve their EV infrastructure deployment strategies. By adopting this comprehensive accessibility measure, other regions can more effectively plan their charging station networks to support the growing number of EV users and encourage the adoption of environmentally friendly transportation options.

Finally, this study calls for ongoing research to refine and validate the proposed method across different urban contexts and to explore its applicability to other types of infrastructure beyond EV charging stations. Future studies could also incorporate data sources like Mobile Device Location Data to capture real-time trip patterns and origins and destinations (51,52). Such research is crucial for developing more sustainable urban environments and for supporting the transition to renewable energy sources in transportation.

In conclusion, the shift in methodology presented in this study represents a significant advancement in how we understand and plan for infrastructure related to EVs. It encourages a move towards more equitable, efficient, and practical solutions in urban planning and infrastructure development, essential for fostering sustainable mobility and addressing the broader environmental challenges of our time.

## 6. References

1. Chapman L. Transport and climate change: a review. Journal of transport geography. 2007;15(5):354–67.

2. Biresselioglu ME, Kaplan MD, Yilmaz BK. Electric mobility in Europe: A comprehensive review of motivators and barriers in decision making processes. Transportation Research Part A: Policy and Practice. 2018;109:1–13.

3. Bakker S, Trip JJ. Policy options to support the adoption of electric vehicles in the urban environment. Transportation Research Part D: Transport and Environment. 2013;25:18–23.

4. Egbue O, Long S. Barriers to widespread adoption of electric vehicles: An analysis of consumer attitudes and perceptions. Energy policy. 2012;48:717–29.

5. She ZY, Sun Q, Ma JJ, Xie BC. What are the barriers to widespread adoption of battery electric vehicles? A survey of public perception in Tianjin, China. Transport Policy. 2017;56:29–40.

6. Baik Y, Hensley R, Hertzke P, Knupfer S. Making electric vehicles profitable. McKinsey & Company. 2019;

7. Lutsey N, Nicholas M. Update on electric vehicle costs in the United States through 2030. Int Counc Clean Transp. 2019;12:1–12.

8. Halbey J, Kowalewski S, Ziefle M. Going on a road-trip with my electric car: Acceptance criteria for long-distance-use of electric vehicles. In Springer; 2015. p. 473–84.

9. Axsen J, Kurani KS. Who can recharge a plug-in electric vehicle at home? Transportation Research Part D: Transport and Environment. 2012 Jul 1;17(5):349–53.

10. Funke SÁ, Sprei F, Gnann T, Plötz P. How much charging infrastructure do electric vehicles need? A review of the evidence and international comparison. Transportation Research Part D: Transport and Environment. 2019 Dec 1;77:224–42.

11. Zou T, Khaloei M, MacKenzie D. Effects of charging infrastructure characteristics on electric vehicle preferences of new and used car buyers in the United States. Transportation Research Record. 2020;2674(12):165–75.

12. Hsu CW, Fingerman K. Public electric vehicle charger access disparities across race and income in California. Transport Policy. 2021;100:59–67.

13. Khan HAU, Price S, Avraam C, Dvorkin Y. Inequitable access to EV charging infrastructure. The Electricity Journal. 2022;35(3):107096.

14. Li J, Wang G, Wang X, Du Y. Smart charging strategy for electric vehicles based on marginal carbon emission factors and time-of-use price. Sustainable Cities and Society. 2023;96:104708.

15. Palaniyappan B, R SK, T V. Dynamic pricing for load shifting: Reducing electric vehicle charging impacts on the grid through machine learning-based demand response. Sustainable Cities and Society. 2024;103:105256.

16. La Rosa D, Takatori C, Shimizu H, Privitera R. A planning framework to evaluate demands and preferences by different social groups for accessibility to urban greenspaces. Sustainable Cities and Society. 2018 Jan 1;36:346–62.

17. Tahmasbi B, Mansourianfar MH, Haghshenas H, Kim I. Multimodal accessibility-based equity assessment of urban public facilities distribution. Sustainable Cities and Society. 2019 Aug 1;49:101633.

18. Miller EJ. Accessibility: measurement and application in transportation planning. Transport Reviews. 2018;38(5):551–5.

19. Hansen WG. How accessibility shapes land use. Journal of the American Institute of planners. 1959;25(2):73–6.

20. Lee G, Lim H. A spatial statistical approach to identifying areas with poor access to grocery foods in the city of Buffalo, New York. Urban Studies. 2009;46(7):1299–315.

21. Chen Y, Bouferguene A, Shen Y, Al-Hussein M. Assessing accessibility-based service effectiveness (ABSEV) and social equity for urban bus transit: A sustainability perspective. Sustainable Cities and Society. 2019 Jan 1;44:499–510.

22. Kelobonye K, McCarney G, Xia J (Cecilia), Swapan MSH, Mao F, Zhou H. Relative accessibility analysis for key land uses: A spatial equity perspective. Journal of Transport Geography. 2019 Feb 1;75:82–93.

23. Esmaili A, Oshanreh MM, Naderian S, MacKenzie D, Chen C. Assessing the spatial distributions of public electric vehicle charging stations with emphasis on equity considerations in King County, Washington. Sustainable Cities and Society. 2024 Jul 15;107:105409.

24. Falchetta G, Noussan M. Electric vehicle charging network in Europe: An accessibility and deployment trends analysis. Transportation Research Part D: Transport and Environment. 2021 May 1;94:102813.

25. Li G, Luo T, Song Y. Spatial equity analysis of urban public services for electric vehicle charging—Implications of Chinese cities. Sustainable Cities and Society. 2022;76:103519.

26. Carlton GJ, Sultana S. Electric vehicle charging station accessibility and land use clustering: A case study of the Chicago region. Journal of Urban Mobility. 2022;2:100019.

27. Chen Y, Chen Y, Lu Y. Spatial Accessibility of Public Electric Vehicle Charging Services in China. ISPRS International Journal of Geo-Information. 2023;12(12):478.

28. Peng Z, Wang MWH, Yang X, Chen A, Zhuge C. An analytical framework for assessing equitable access to public electric vehicle chargers. Transportation Research Part D: Transport and Environment. 2024 Jan 1;126:103990.

29. Bhat C, Handy S, Kockelman K, Mahmassani H, Chen Q, Srour I, et al. Assessment of accessibility measures. Work. 2001;7:4938–3.

30. El-Geneidy AM, Levinson DM. Access to destinations: Development of accessibility measures. 2006;

31. Talen E, Anselin L. Assessing spatial equity: an evaluation of measures of accessibility to public playgrounds. Environment and planning A. 1998;30(4):595–613.

32. Potoglou D, Song R, Santos G. Public charging choices of electric vehicle users: A review and conceptual framework. Transportation Research Part D: Transport and Environment. 2023;121:103824.

33. Latinopoulos C, Sivakumar A, Polak JW. Response of electric vehicle drivers to dynamic pricing of parking and charging services: Risky choice in early reservations. Transportation Research Part C: Emerging Technologies. 2017 Jul 1;80:175–89.

34. Ma SC, Yi BW, Fan Y. Research on the valley-filling pricing for EV charging considering renewable power generation. Energy Economics. 2022 Feb 1;106:105781.

35. Wang Y, Yao E, Pan L. Electric vehicle drivers' charging behavior analysis considering heterogeneity and satisfaction. Journal of Cleaner Production. 2021;286:124982.

36. Yong JY, Tan WS, Khorasany M, Razzaghi R. Electric vehicles destination charging: An overview of charging tariffs, business models and coordination strategies. Renewable and Sustainable Energy Reviews. 2023;184:113534.

37. Ge Y, MacKenzie D. Charging behavior modeling of battery electric vehicle drivers on long-distance trips. Transportation Research Part D: Transport and Environment. 2022;113:103490.

38. Li H, Yu L, Chen Y, Tu H, Zhang J. Uncertainty of available range in explaining the charging choice behavior of BEV users. Transportation Research Part A: Policy and Practice. 2023 Apr 1;170:103624.

39. Mehditabrizi A, Namadi SS, Cirillo C. Tackling Ev Adoption Challenges: Insights from Refueling Behavior Analysis. 2024.

40. Sun XH, Yamamoto T, Morikawa T. Fast-charging station choice behavior among battery electric vehicle users. Transportation Research Part D: Transport and Environment. 2016 Jul 1;46:26–39.

41. Visaria AA, Jensen AF, Thorhauge M, Mabit SE. User preferences for EV charging, pricing schemes, and charging infrastructure. Transportation Research Part A: Policy and Practice. 2022 Nov 1;165:120–43.

42. Team ChargePoint. What's the difference between Level 2 AC charging and DC fast charging? [Internet]. 2023. Available from: https://www.chargepoint.com/blog/whats-difference-between-level-2-ac-charging-and-dc-fast-charging

43. Bhat C, Handy S, Kockelman K, Mahmassani H, Chen Q, Weston L. Urban accessibility index: literature review. Center of Transportation Research, University of Texas at Austin, Springfield. 2000;

44. Chang Z, Chen J, Li W, Li X. Public transportation and the spatial inequality of urban park accessibility: New evidence from Hong Kong. Transportation Research Part D: Transport and Environment. 2019;76:111–22.

45. Geurs KT, Ritsema van Eck JR. Accessibility measures: review and applications. Evaluation of accessibility impacts of land-use transportation scenarios, and related social and economic impact. RIVM rapport 408505006. 2001;

46. Bunel M, Tovar E. Key issues in local job accessibility measurement: Different models mean different results. Urban Studies. 2014;51(6):1322–38.

47. Moran PA. Notes on continuous stochastic phenomena. Biometrika. 1950;37(1/2):17–23.

48. Dale MR, Fortin MJ. Spatial analysis: a guide for ecologists. Cambridge University Press; 2014.

49. Saleh Namadi S, Tahmasbi B, Mehditabrizi A, Darzi A, Niemeier D. Using Geographically Weighted Models to Explore Temporal and Spatial Varying Impacts on Commute Trip Change Resulting from COVID-19. Transportation Research Record. 2024;03611981241231797.

50. Anselin L. Local indicators of spatial association—LISA. Geographical analysis. 1995;27(2):93–115.

51. Kabiri A, Darzi A, Pan Y, Saleh Namadi S, Zhao G, Sun Q, et al. Elaborated Framework for Duplicate Device Detection from Multisourced Mobile Device Location Data. Transportation Research Record. 2024;2678(6):881–90.

52. Yang M, Luo W, Ashoori M, Mahmoudi J, Xiong C, Lu J, et al. Big-data driven framework to estimate vehicle volume based on mobile device location data. Transportation research record. 2024;2678(2):352–65.